\documentstyle[12pt]{article}
\newcommand{\be}{\begin{equation}}
\newcommand{\ee}{\end{equation}}

\newcommand{\bt} { \begin{tabular} }
\newcommand{\et}{ \end{tabular} }
\newcommand{\bc} { \begin{center} }
\newcommand{\ec}{ \end{center} }

\newcommand{\f}{ \frac }

\newcommand{\la}{\label }
\newcommand{\bfi}{\begin{figure} }
\newcommand{\efi}{\end{figure} }

\newcommand{\btb} { \begin{table} }
\newcommand{\etb}{ \end{table} }
\newcommand{\mn} {\matrix{ _{_{>}} \cr ^{^{<}} \cr} }
\newcommand{\nm} {\matrix{ _{_{<}} \cr ^{^{>}} \cr} }
\newcommand{\ns} { \matrix{ _{_{<}} \cr ^{^{\sim}} \cr} }
\begin{document}
\title{Critical and off-critical studies of the Baxter-Wu model
with general toroidal boundary conditions}

\author{F. C. Alcaraz \ \ and \ \  J. C. Xavier \\
	 Departamento de F\'\i sica \\
	Universidade Federal de S\~ao Carlos \\
	 13565-905, S\~ao Carlos, SP, Brasil }
\date{PACS numbers : 05.50+q,64.60Cn, 75.10Jn }
\maketitle
\vspace{0.2cm}
\begin{abstract}
The operator content of the Baxter-Wu model with general toroidal 
boundary conditions is calculated analytically and numerically. 
These calculations were done by relating the partition function of the 
model with the generating function of a site-colouring problem in a 
hexagonal lattice. Extending the original Bethe-ansatz solution of the 
related  colouring problem we are able to calculate the eigenspectra of 
both models by solving the associated Bethe-ansatz equations. We have also 
calculated, by exploring the conformal invariance at the critical point, 
the mass ratios of the underlying massive theory governing the Baxter-Wu 
model in the vicinity of its critical point.
\end{abstract}
\section{Introduction}
The Baxter-Wu model is the simplest non-trivial spin model with three-spin 
interactions. Its Hamiltonian is defined on a triangular lattice by 
\be
H=-J\sum_{<ijk>}\sigma_i\sigma_j\sigma_k ,
\la{ham}
\ee
where the sum extends  over the elementary triangles, $J$ is the coupling constant   
and $\sigma_i=\pm 1$ are Ising variables located at the sites.
Historically this model was introduced by Wood and Griffiths in 1972 \cite{wg}, 
 as an example of a model exhibiting an order-disorder phase transition and not 
having a global up-down spin reversal symmetry. 
This model is  self-dual \cite{wg,merlini}
with the same critical temperature as that of  the Ising model on a 
square lattice. In 1973 Baxter and Wu \cite{bw} related the partition function 
of this model, in the   thermodynamic limit, with the generating function
 of a site colouring problem on an hexagonal lattice. Solving this colouring
 problem through a generalized Bethe-ansatz they calculated the leading
 exponents \cite{bw} $\alpha =2/3$, $\mu=2/3$ and  $\eta=1/4$ of the 
Baxter-Wu model.
The equality of these exponents with those of the  4-state Potts model
 [4-6] added to the fact that both models have the same four-fold degeneracy
 of the ground state, induce the conjecture that they share the same
 universality class of critical behaviour. This conjecture is 
interesting since contrary to the Baxter-Wu model the 4-state Potts model 
is exactly integrable only at its critical point. However from  numerical 
 studies  of these  models on a finite lattice it is well known that 
both models show different corrections to finite-size scaling 
(at the critical point). Whereas in the 4-state Potts models [7-10]
these corrections are governed by a marginal operator, producing
logarithmic corrections with the system size, which brings enormous 
difficulty to extract reliable results from finite-size studies of the 
model, this is not the case in the Baxter-Wu model [11-13].   

Nowadays with the developments of conformal invariance applied to  critical
 phenomena  \cite{cardrev} the classification of different
 universality classes of critical behaviour  becomes clear. Two models 
belong to the same universality class only if they share the same operator
 content, not only the leading critical exponents. The operator content of 
the 4-state Potts model was already conjectured from finite-size studies
 in its Hamiltonian formulation
\cite{baake} and can be obtained by a $Z(2)$ orbifold of the
Gaussian model in a special compactification  radius (see \cite{ginsparg} 
for a review). In this paper we numerically and analytically calculate the 
operator content of the Baxter-Wu model with several boundary conditions. 
Part of our numerical calculation was announced in 
\cite{ax}. In order to do this calculation we generalize the original
 Bethe-ansatz solution of the related site-colouring problem with periodic
 boundary condition. This is necessary because this solution only gives 
part of the eigenspectrum of the associated transfer matrix. We also 
extend our Bethe-ansatz solution in other cases where the site-colouring
 problem is not on a periodic lattice. This extension enables us to obtain 
the operator content of this model and the Baxter-Wu model for more general
toroidal boundary  conditions. Our numerical study was done by 
numerically solving the Bethe-ansatz equations and the analytical work was done 
by studying  these equations  using standard techniques based on the
Wiener-Hopf method \cite{morse}.

We believe that beyond the Ising model the Baxter-Wu model is the simplest 
spin model that can be solved exactly for arbitrary temperatures. We 
explore this solution in order to obtain the mass spectrum of the massive 
theory describing fluctuations near the critical point.

The layout of this paper is as follows. In section 2 we introduce a 
site-colouring problem on an hexagonal lattice whose generating function 
is exactly related, in the  bulk limit, with the Baxter-Wu model. 
Our construction is valid for some toroidal boundary conditions, 
generalizing the original construction \cite{bw} for periodic lattices. 
In section 3 we present the transfer matrix associated to the colouring 
problem and calculate its eigenvalues by the Bethe-ansatz. In section 4 
the operator content of the Baxter-Wu model and the related site-colouring
 problem is obtained. The mass spectrum of the massive field theory governing
the thermal and magnetic perturbations is calculated in section 5. Finally our
 conclusions are present in section 6 and the analytical calculation of some 
of the conformal dimensions of the site-colouring problem is presented in 
an appendix.
\section{The Baxter-Wu model and the related site-colouring problem}
In this section we relate the Baxter-Wu model with general toroidal boundary 
conditions with a {\it site-colouring problem} ($SCP$) on the hexagonal lattice. 
The construction presented here generalizes those presented by Baxter and
Wu \cite{bw} for the periodic case.

Let us consider a triangular lattice with L (M) rows (columns) along the
 horizontal (vertical) direction, respectively. For convenience we take $L$ as
a multiple of 3, and decompose the lattice in three triangular sublattices 
formed by the points denoted by {\Large$\circ $}, {\large{$\Box $}}
   and $\triangle $ in Fig. 1. We attach at each site of   a sublattice Ising
 variables $\{ \sigma^{ \circ } \}$, $\{ \sigma^{ \Box  } \}$ and 
$\{ \sigma^{ \triangle  } \}$, and the Baxter-Wu model, with the Hamiltonian 
 (\ref{ham})  is given in terms of the simplest three-body interactions with 
one spin in each sublattice.

The model has a non-local $Z(2)\times Z(2)$ symmetry corresponding to the global 
change of variables in two of the sublattices, i.e.,
$$
\begin{array}{llll}
\sigma^{ \triangle  }\rightarrow a \sigma^{ \triangle } &
\sigma^{ \circ  }\rightarrow b \sigma^{ \circ  } & 
\sigma^{ \Box  }\rightarrow ab \sigma^{ \Box  } &  (a=b=\pm 1). \\
\end{array}
$$
This fact implies that at sufficiently low temperatures the model have a 
four-fold degenerate ground state. Creating, at low temperatures,
 domain walls connecting those different ground states we see that in  fact 
the symmetry of those long-range excitations is $D(4)$, like in the 4-state
 Potts model. These reasonings suggest that critical fluctuations of both 
models are described in terms of the same quantum field theory. In order to 
do a detailed study of several sectors of this underlying field theory we
 consider the Baxter-Wu model with general toroidal boundary conditions
 compatible with its $Z(2)\times Z(2)$ symmetry:
\be
\begin{array}{lll}
\sigma^{ \triangle  }_{i,j}= a \sigma^{ \triangle }_{i+L,j} &
\sigma^{ \circ  }_{i,j}= b \sigma^{ \circ  }_{i+L,j} & 
\sigma^{ \Box  }_{i,j}= ab \sigma^{ \Box  }_{i+L,j}, \\
\end{array}
\la{sigma}
\ee
where $a,b=\pm 1$. In Fig.1 the simbols 
{\Large$\circ $}, {\large{$\Box $}} and $\triangle $ denote the bordering sites
which are related by (2) to the bulk ones (filled simbols).
 Imposing periodic boundary condition along the vertical direction the partition 
function of the model $Z_{L\times M}^{BW}$ can be written as
\be
Z_{L\times M}^{BW}=Tr\left( \hat T_{(a,b)}^{BW}\right)^M,
\la{pbw}
\ee
where $\hat T_{(a,b)}^{BW}$ is the associated row-to-row transfer \ matrix. 
Its \, \,\, elements 
$\langle\sigma_1,...,\sigma_N|\hat T_{(a,b)}^{BW}|\sigma_1',...,\sigma_N'\rangle$
are given by the Boltzmann weights generated by the spin configurations
$\{\sigma_1,...,\sigma_N\}$ and $\{\sigma_1',...,\sigma_N'\}$ of adjacent rows
and are given by
\be
\langle\sigma_1,...,\sigma_N|\hat T_{(a,b)}^{BW}|\sigma_1',...,\sigma_N'\rangle
=\exp\left(
K\sum_{j=1}^{L} \sigma_j'\sigma_{j+2}(\sigma_{j+1}+\sigma_{j+1}')\right),
\la{peso}
\ee
with $K=\f{J}{k_BT}$ and on the right side the appropriate boundary condition $(a,b)$ 
is taken into account.

Following Baxter and Wu \cite{bw} we can relate $Z_{L\times M}^{BW}$ to the
partition function or generating function $Z_{N\times M}^{SCP}$ of a
$SCP$ on an honeycomb lattice with $N=\f{2L}{3}$  rows and $M$ columns. In 
order to do this, it is convenient to  define link variables $\{\lambda\}$ 
at the links of the hexagonal lattice formed by the sublattices {\Large$\circ$} 
and $\Box$ (heavy lines in Fig.1). These variables are given by the product of
 the site variables at the ends of the link ($\lambda=\sigma^\Box\sigma^{\circ}$). 
In terms of these new variables the Hamiltonian is given by
\be
H=-J\sum_{i\in\triangle}
\sigma_i^{\triangle}(\lambda_1+\lambda_2\cdots+\lambda_6),
\la{hnew}
\ee
where the summation is only over the sublattice $\triangle$ and 
$\lambda_1,\lambda_2,...,\lambda_6$ are the link variables surrounding a
given site variable $\sigma_i^\triangle$. Taking into account the boundary 
 conditions (\ref{sigma}) the partition function, in terms of these new variables, 
can be written as 
{\small
\begin{equation}
Z_{L\times M}^{BW}=\sum_{\left\{ \sigma^{\triangle},\lambda \right\} }
\prod_{\triangle}\left\{
\exp \left\{ K\sigma^{\triangle}(f_1\lambda
_1+f_2\lambda _2+\cdot \cdot \cdot +f_6\lambda _6)\right\} \frac
12(1+\lambda _1\lambda _2...\lambda _6)\right\},  
\label{z1}
\end{equation}
}
where the product extends over all the elementary hexagons formed by the 
sites in sublattices $\Box$ and {\Large$\circ$},
 and surrounding  a site variable 
$\sigma^{\triangle}$. The factor 
$\f12(1+\lambda _1\lambda _2...\lambda _6)$ in (\ref{z1}) is necessary 
since the variables $\{\lambda\}$ are not independent. The factors $f_i$
 ($i=1,2,...,6$) in (\ref{z1}) are constants defined on the links of the
 hexagonal lattice $(\Box$-{\Large$\circ$}) and depend on their relative 
location (see Fig. 1). If both ends of a link $i$ (with corresponding 
variable $\lambda_i$) belongs to the bulk of the lattice 
($\bullet$-\rule{2mm}{2mm})  $f_i=1$, if one of its ends is on the
border ({\Large$\circ$}-\rule{2mm}{2mm} or $\bullet-\Box$) $f_i=ab$, and if
 both ends are on the  border ({\Large$\circ$}-$\Box$) $f_i=a$. To proceed it 
is  convenient \cite{bw} to rewrite the last product in (\ref{z1}), 
surrounding each lattice site $\triangle$ as
\be
\frac 12(1+\lambda _1\lambda _2...\lambda _6)=
\sum_{\mu^\triangle =\pm1} g(\lambda _1,\mu^\triangle )g(\lambda
_2,\mu^\triangle )\cdot \cdot \cdot g(\lambda _6,\mu^\triangle ),
\la{suml} 
\ee
where
\be
g(\lambda,\mu)=2^{-7/6}(\lambda+\mu+|\lambda-\mu|).
\la{gs}
\ee
Substituting the expression (\ref{suml}) into equation (\ref{z1})
 the partition function 
will be expressed in terms of a single sublattice $\triangle$ with site 
variables $\{\sigma^\triangle\}$ and $\{\mu^\triangle\}$, and link variables
$\{\lambda\}$. Taking into account the boundary factors it is straightforward 
to write
\be
Z_{L\times M}^{BW}=\sum_{ \{ \sigma^\triangle,\mu^\triangle \} }
\prod_{< i,j >}
\left\{
\sum_{\lambda_{i,j} =\pm 1}\exp
 \left( K(\sigma_i^\triangle+\sigma_j^\triangle)\lambda_{i,j} \right)
g(\lambda_{i,j} ,\mu_i^\triangle )g(\lambda_{i,j} ,\mu_j^\triangle ) \right\},
\la{z2}
\ee
where $< i,j >$ are links on the hexagonal lattice formed by the sublattices 
$\Box$ and {\Large$\circ$} (see Fig. 1). In order to derive (\ref{z2}) we are forced to
restrict ourselves only to boundary  conditions (\ref{sigma}) where $a=b=\pm1$.
Fortunately this does not restrict our analysis since due to the $D(4)$ symmetry
of the model, the eigenspectra of the  transfer matrices
 with boundary  condition 
$a\neq b$ and $a=b=-1$ are  degenerate. Summing over $\{\lambda\}$ in 
(\ref{z2}) we obtain
\be
Z_{L\times M}^{BW}=\sum_{ \{ \sigma^\triangle, \mu^\triangle \} }
\prod_{< i,j > }
w(\sigma_i^\triangle,\mu_i^\triangle ;
\sigma_j^\triangle,\mu_j^\triangle ),
 \label{z3}
\ee
where 
\be
w(\sigma_i,\mu_i ;
\sigma_j,\mu_j )
=
2^{-1/3}\left[ \exp \left(K(\sigma_i+\sigma_j)\right)
+\mu_i \mu_j\exp \left(-K(\sigma_i+\sigma_j)\right)\right].
\la{w2}
\ee

Finally, following Baxter and Wu \cite{bw} we now associate the 
above partition function to the generating function of a $SCP$ with
 8 colours. We associate odd colours 1, 3, 5 and 7 at a given site 
$\triangle$ according to the values of the variables 
$(\sigma^\triangle,\mu^\triangle)$ attached at the site:
\be
\begin{array}{llll}
(++) \rightarrow 1 & 
(-+) \rightarrow 3 & 
(--) \rightarrow 5 & 
(++) \rightarrow 7 . \\
\end{array}
\la{ncor}
\ee
The relation (\ref{w2}) tell us that links connecting colours 3,7 
and 1,5 should be forbidden in the $SCP$. This constraint can be easily 
implemented \cite{bw} by introducing the even colours 2, 4, 6 and 8 on
 the sublattice {\Large$\circ$}, 
forming with the sublattice $\triangle$ an hexagonal
lattice with $N=\f{2}{3}L$ $(M)$ sites in the horizontal (vertical) 
direction. The constraint is that all nearest-neighbour colours on this
hexagonal lattice must differ by $\pm 1$ (modulo 8). For
 $M \rightarrow \infty$, the partition function $Z_{L,M}^{BW}$ is
then related to the generating function $Z_{N,M}^{SCP}$, given by
\be
Z_{N,M}^{SCP}=
\sum_{\{C\}} z_1^{n_1}z_2^{n_2}\cdot \cdot \cdot z_8^{n_8}
=Z_{L,M}^{BW},
\la{pbwscp}
\ee
where $n_1,n_2,...,n_8$ is the number of sites coloured with colour 
1,2,...,8 in a given configuration $C$. The fugacities $z_i$ 
$(i=1,2,...,8)$ are obtained from (\ref{w2})
\be
\begin{array}{l}
z_1=z_3=z_5=z_7= 2\sinh ( 4K), \\
z_2^{-1}=z_4=z_6^{-1}=z_8=\sinh (2K)\equiv t. 
\end{array} 
\la{t}
\ee
The critical point of the Baxter-Wu model and of the $SCP$ is given by 
the self-dual point  $t=t_c=1$ \cite{bw}. Concerning the boundary conditions
the relations are as follows. The $Z_{L,M}^{BW}$ with periodic boundary
condition $(a=b=1$ in (\ref{sigma})) is related with $Z_{N,M}^{SCP}$, also
with  periodic lattice. The boundary condition 
$a=b=-1$ in (2) is related to a boundary  condition in the $SCP$ such that
if at site $(1,j)$ we have a colour variable $c_{1,j}=1,3,5$ or 7 at
site $(N+1,j)$ the colour variable should be $c_{N+1,j}=3,1,7$ and 5
respectively. We do not need to mention even colours since
$N$ is even (see Fig. 1).
\section{The transfer matrix and the Bethe-ansatz of the $SCP$}
In this section we derive the row-to-row transfer matrix of the $SCP$ and
 generalize its Bethe-ansatz solution presented by Baxter and Wu 
\cite{bw}. The solution presented in \cite{bw}, as we shall see, only 
gives part of the eigenspectra of the transfer matrix of the $SCP$, 
with periodic boundary conditions. Here we extend the solutions for 
the periodic case and also derive the Bethe-ansatz equations for the 
more general boundary  condition
\be
c_{i,j}=c_{i+N,j}+2\kappa \hspace{.2cm}(\mbox{mod} \hspace{.2cm}8);
\hspace{.5cm}\kappa=0,1,2\hspace{.2cm} \mbox{or}\hspace{.2cm} 3,
\la{cor}
\ee
along the horizontal direction. The periodic case corresponds to $\kappa=0$.
Following \cite{bw}, for a given configuration $\{c_{i,j}\}$ of colours on
the hexagonal lattice, we say we have a dislocation on a given link of the 
lattice wherever the colour on the right end of the link is smaller 
(modulo 8) than that on the left end. In Fig. 2 we show two  configurations 
with the corresponding dislocations (dotted lines) 
for a lattice with width $N=4$. In Fig. 2a the lattice is periodic 
($\kappa=0$) and in Fig. 2b the boundary condition is given by (\ref{cor}) 
with $\kappa=1$.

The generating function of the $SCP$ can be written as
\be
Z_{N,M}^{SCP}=Tr\left( \hat T_{(\kappa)}^{SCP}\right)^M ,
\la{pscp}
\ee
where $\hat T_{(\kappa)}^{SCP}$ is the associated row-to-row transfer
matrix. This  transfer matrix has elements 
$\langle \{C\}|\hat T_{(\kappa)}^{SCP}|\{C'\} \rangle$ given by the product 
of the Boltzmann 
weights due to the colour  configurations $\{C\}=\{c_1,c_2,..,c_N\}$ and
$\{C'\}=\{c_1',c_2',...,c_N'\}$ of two adjacent rows. If the configuration
 produced by $\{C\}$ and $\{C'\}$ contains only colours differing by
$\pm 1$ (modulo 8), we have
\be
\langle \{C\}|\hat T_{(\kappa)}^{SCP}|\{C'\} \rangle =
\prod_{i=1}^{N}(z(c_i)z(c_i^{\prime}))^{1/2},
\la{mscp}
\ee
with fugacities $z_i$  defined in (\ref{t}). On the other hand if the
 configuration does not satisfy this constraint,
$\langle \{C\}|\hat T_{(\kappa)}^{SCP}|\{C'\} \rangle =0$. It is simple to see 
that this 
requirement implies, for arbitrary values of $\kappa$ in (\ref{cor}), the 
conservation of the number $n$ of dislocations  along the vertical direction. 
Consequently the Hilbert space associated to $\hat T_{(\kappa)}^{SCP}$ 
can be separated into block-disjoint sectors labelled by the values
of $n$. The possible values of $n$ depend on the boundary condition
(\ref{cor}). For $\kappa=0,$ 2 (1, 3) they are even (odd), and are given
by $n_j$, where 
\be
0\le n_j=N- \kappa-4j\le 2N
\hspace{.5cm}j=0,\pm1\pm2,...
\la{dis}
\ee 

The colour configurations $\{C\}$ and $\{C'\}$, in a sector with $n$ dislocations, 
can be conveniently expressed by the sets 
$(m;X)=(m;x_1,x_2,...,x_n)$ and $(m';X')=(m';x_1',x_2',...,x_n')$, 
respectively. The odd numbers $m$ and $m'$ give the colour at the
first site and the sets $X\equiv(x_1,x_2,...,x_n)$ and 
 $X'\equiv(x_1',x_2',...,x_n')$ give the position of the 
dislocations on the row. The sets $X$ and $X'$ should satisfy
\be
1\leq x_1\leq x_2\leq\cdot \cdot \cdot \leq x_n\leq N 
\hspace{.25cm},\hspace{.25cm}
1\leq x_1'\leq x_2'\leq\cdot \cdot \cdot \leq x_n'\leq N,
\la{xy0}
\ee 
and should have no more than one  repeated value of
$x$ or $x'$ if they are odd, or more than three repeated values 
if they are even.  In Fig. 2a we show the configurations 
$(m;X)=(1;2,3,4,4)$ and $(m';X')=(3;2,2,3,4)$ which belong to the
sector with $n=4$ and periodic boundary  condition. In Fig 2b we
show the configurations $(m;X)=(1;2,3,4)$ and $(m';X')=(3;2,2,2)$ 
belonging to the sector with $n=3$ in the lattice with boundary 
condition $\kappa=1$ in (\ref{cor}) and width $N=4$.
Two configurations $X=(x_1,x_2,...,x_n)$ and $X'=(x_1',x_2',...,x_n')$
are connected through the operator $\hat T_{(\kappa)}^{SCP}$ if
beyond (\ref{xy0})
\be
\begin{array}{lll}
x_j' &=&x_j-1,\hspace{2cm}  
\mbox{if} \hspace{.25cm} x_j \hspace{.25cm} \mbox{odd} \\ 
x_j' &=&x_j,x_j-1,x_j-2\,\,\,\,\,\mbox{if} \hspace{.25cm} x_j\,\,\, \mbox{even.}
\la{xy}
\end{array}
\ee
In the case $x_i=x_{i+1}$ we have an additional constraint
$x_i'\neq x_{i+1}'$. We should notice, for arbitrary boundary condition, the identification
\be
X=(0,x_2,x_3,...,x_n)=(x_2,x_3,...,x_n,N).
\la{xcon}
\ee
The transfer matrix, in a given sector with $n$ dislocations is now given by
{\small
\be
\langle m,X|\hat T_{(\kappa)}^{SCP}|m',X' \rangle =z_{m+1}^{ \f{1-(-1)^{\kappa} }{4} }
\prod_{j=1}^nw(m+x_j+x_j'-2j),\hspace{.2cm} w(m)=(z_mz_{m+1})^{1/2}, 
\label{tc}
\ee
}
if $X$ and $X'$ satisfy (\ref{xy0}) and
(\ref{xy}), and is zero otherwise. In the sector with $n$ dislocations the
eigenvectors $\psi^{(n)}$ of $\hat T_{(\kappa)}^{SCP}$, with eigenvalue
 $\Lambda$ can be written as 
\be
\psi^{(n)}=\sum_{ \{m,X\} }z_{m+1}^{ -\f{1-(-1)^{\kappa} }{8} }
f_m(x_1,x_2,\cdot \cdot \cdot ,x_n)|m;X \rangle, 
\la{wfunction}
\ee
where the summation is restricted to the configurations with $n$ dislocations, 
and $f_m(x_1,x_2,\cdot \cdot \cdot ,x_n)$ are unknown
amplitudes. The eigenvalue equation for $\hat T_{(\kappa)}^{SCP}$ is given by
\be
\sum_{X'}^{*} \left( \prod_{j=1}^n w(m+x_j+x_j'-2j)\right) f_{m+2}(X')=
\Lambda
f_m(X) ,  
\label{eqauto}
\ee
where the asterisk indicates that $X$ and $X'$ satisfy the conditions
(\ref{xy0}) and (\ref{xy}). The relation (\ref{xcon}) implies that
in (\ref{eqauto}) the amplitudes having $x_1'=0$ should be replaced
by the boundary  condition
\be
f_{m+2}(0,x_2',x_3',\cdot \cdot \cdot ,x_n')=z_{m+1}^{ \f{1-(-1)^{\kappa}}{4} }
f_m(x_2',x_3',\cdot \cdot \cdot ,x_n',N).
\label{ccontorno}
\ee

Due to the values of the fugacities (\ref{t}) it is simple to see that
$\hat T_{(\kappa)}^{SCP}$, besides conserving the number of dislocations, also
has an additional $Z(2)$ symmetry (eigenvalues $\epsilon=\pm1$), 
since adding 4 (modulo 8) to all colours in a given configuration
does not change its weight in the generating function, that is
\be
f_{m+4}(X)=\epsilon f_m(X).
\la{eps}
\ee

Following Baxter and Wu \cite{bw} we assume the following Bethe-ansatz
for the amplitudes
\be
f_m(X)=\sum_Pa(P)\phi _{P_1}(m-2,x_1)\cdot \cdot \cdot \phi _{P_n}(m-2n,x_n),
\label{ansatz}
\ee
where  the summation is over all the $n!$ permutations 
$P=\{P_1,P_2,...,P_n\}$ of integers $\{1,2,...,n\}$. 
We require the existence of $n$ wave numbers $k_j$ ($j=1,2,...,n$) 
and signs $\epsilon_j$ ($j=1,2,...,n$) such that
\begin{equation}
\phi _j(m,x)=\epsilon _j\phi _j(m+4,x)=\left\{ 
\begin{array}{l}
a_{j,m}\exp (ik_jx), \\ 
b_{j,m}\exp (ik_jx),
\end{array}
\right. 
\begin{array}{l}
\,\,x\,\, \mbox{ odd} \\ 
\,\,x\,\, \mbox{ even.}
\end{array}
\la{oe}
\end{equation} 
Observe that the $Z(2)$-parity eigenvalue of the wave function is given by
$\epsilon=\prod_{j=1}^{n}\epsilon_j$, and it is even or odd depending 
on the numbers of negative values of $\epsilon_j$. The Bethe-ansatz
solution presented by Baxter and Wu \cite{bw} only gives the symmetric
eigenvalues ($\epsilon=1$), for periodic boundary conditions ($\kappa=0$). We
can follow the same procedure as in \cite{bw} in order to derive the 
Bethe-ansatz equations. We have to consider various possible choices of $X$
to determine the eigenvalue $\Lambda$.

Firstly let us consider the case where all dislocations are located at
 distinct positions, i.e., $x_1\neq x_2\neq\cdots\neq x_n$. 
Eq. (\ref{eqauto}) is then replaced by
\be
\sum_{x'}w(m+x+x')\phi _j(m+2,x')=\lambda _j\phi _j(m,x),\,\,\,\,
\,\,j=1,...,n , 
\label{d1}
\ee
where we have denoted
\be
\Lambda =\lambda _1\cdot \cdot \cdot \lambda _n.  
\label{lamb2}
\ee
Actually Eq. (\ref{d1}) represents two equations corresponding to 
$x$ odd or even and can be written as
\be
T_{j,m+2}V_{j,m+2}={\lambda}_jV_{j,m},
\label{tv}
\ee
where 
\be
\begin{array}{cc}
T_{j,\,m+2}=\left( 
\begin{array}{cc}
0 & \alpha _{j,\,m+1} \\ 
\alpha _{j,m-1} & A_{j,m}
\end{array}
\right); & V_{j,\,m}=\left( 
\begin{array}{c}
a_{j,m} \\ 
b_{j,m}
\end{array}
\right)
\end{array}
\label{ab}
\ee
with
\be
\alpha _{j,\,m}=w(m)\exp (-ik_j) ;\hspace{.25cm} 
A_{j,\,m}=w(m)+w(m-2)\exp (-2ik_j).
\la{eqqq}
\ee
Using the fact that
$$
\left( 
\begin{array}{c}
a_{j,m-2} \\ 
b_{j,m-2}
\end{array}
\right) =\epsilon _j\left( 
\begin{array}{c}
a_{j,m+2} \\ 
b_{j,m+2}
\end{array}
\right),  
$$
it is simple to see that $\lambda_j$ 
can be obtained from
the eigenvalue equation
$$
(T_{j,m}T_{j,m+2})V_{j,m+2}= \lambda_j^2\epsilon_jV_{j,m+2}.
$$
Solving for the eigenvalues of $T_{j,m}T_{j,m+2}$ we see that
$\lambda_j$ does not depend on the value of $m$ (as it is 
expected) and satisfies 
\be
 \lambda_j^4-\epsilon_j\lambda_j^2
[\exp(-i4k_j)+\Delta\exp(-2ik_j)+1]+\exp(-4ik_j) =0,
\la{eqq}
\ee
where $\Delta=t+1/t$. The solution of (\ref{eqq}) is given
by $\lambda_j=\sqrt{\epsilon_j}\exp(e_j^{(s_j)}-ik_j^{(s_j)})$, with
\be
 e_j^{(s_j)}=1/2\ln\left( x_j+s_j\sqrt{x_j^2-1}\right) , \; \; 
x_j=\cos(2k_j^{(s_j)})+\Delta ,
\la{es}
\ee
and $s_j=\pm1$. Equation (\ref{tv}) gives the relations
%
%
\begin{eqnarray}\nonumber
a_{j,m+2}=\epsilon_j\alpha_{j,m-1}b_{j,m}/\lambda_j, \\ \nonumber
b_{j,m+2}=\sqrt{\epsilon_j} \Omega _{j,m}b_{j,m}, \\
a_{j,m}=\sqrt{\epsilon_j}\alpha_{j,m+1}\Omega _{j,m}b_{j,m}/\lambda_j
\la{rela},
\end{eqnarray}
%
%
where 
\be
\Omega _{j,m}=\epsilon_j\sqrt{\epsilon_j}A_{j,m}^{-1}
(\lambda _j-\epsilon_j\alpha _{j\,m-1}^2/\lambda_j^{-1}).
\la{omega}
\ee
It is important to verify that
$\Omega _{j,m+2}\Omega _{j,m}=1$, so that at $t=t_c=1$, 
$\Omega _{j,m}^2=1$.

Secondly let us consider the case where two even dislocation positions
$x$ coincide. For convenience suppose $x_1=x_2=x$ (even). If
we require that the ansatz (\ref{ansatz}) satisfies (\ref{eqauto}) with
eigenvalue given by (\ref{lamb2}) and (\ref{es}) the equation
\be
\sum_{x_1',x_2'}^{*}w(m+x+x_1'-2)w(m+x+x_2'-4)f_{m+2}(x_1',x_2')=
\lambda _1\lambda_2f_m(x,x),  
\label{d2}
\ee
must be fulfilled, where
\be
f_{m+2}(x_1,x_2)=a(1,2)\phi _1(m,x_1)\phi _2(m-2,x_2)+a(2,1)\phi
_2(m,x_1)\phi _1(m-2,x_2),  
\label{fd2}
\ee
and the asterisk in Eq. (\ref{d2}) indicates summation over the 
possible configurations $(x_1',x_2')=(x-2,x-1),(x-2,x),(x-1,x)$.
Using (\ref{oe}), (\ref{ab}), (\ref{eqq}) and (\ref{rela})   in
(\ref{d2}) we obtain after some algebra (the same algebra 
as in \cite{bw}), that the ratio 
\be
\tilde B_{12}\equiv \f{a(1,2)}{a(2,1)}=
-\frac{\sqrt{\epsilon_2}}{\sqrt{\epsilon _1}}
\frac{\cosh (e_1^{(s_1)}+ik_2)}{\cosh (e_2^{(s_2)}+ik_1)}=
\frac{\sqrt{\epsilon _2}}{\sqrt{\epsilon _1}}B_{12},
\la{razao}
\ee
is also independent of $m$

More generally in order for the Bethe-ansatz (\ref{ansatz}) to work
we should have
\be
\tilde B_{jl}\equiv \frac{a(...,j,l,...)}{a(...,l,j,...)}=
-\frac{\sqrt{\epsilon _l}}{\sqrt{\epsilon _j}}
\frac{\cosh (e_j^{(s_j)}+ik_l)}{\cosh (e_l^{(s_l)}+ik_j)}=
\frac{\sqrt{\epsilon _l}}{\sqrt{\epsilon _j}}B_{jl}, 
\label{bb}
\ee
for all permutations of adjacent elements in $a(P)$. Following 
the same steps as in \cite{bw} it can be proved that equations 
(\ref{bb}) are enough to ensure the effectiveness of the ansatz 
(\ref{ansatz}) in the case of triple coincidence of
dislocations. In order to complete the solution we still need to
fix the wave numbers $\{k_1,k_2,...,k_n\}$. As usual this is
done
by inserting (\ref{ansatz}) in the boundary  condition (\ref{ccontorno}),
{\small
\be
\begin{array}{l}
\sum a(P_1,...,P_n)\phi _{P1}(m,0)
\left[ \phi _{P_2}(m-2,x_2)\cdot \cdot \cdot
\phi _{P_n}(m-2n+2,x_n)\right]=z_{m+1}^{ \f{1-(-1)^{\kappa} }{4} }\\
\sum a(P_2,...,P_n,P_1)\left[ \phi _{P_2}(m-2,x_2)\cdot \cdot \cdot \phi
_{P_n}(m-2n+2,x_n)\right] \phi _{P_1}(m-2n,N),
\end{array}
\la{c1} 
\ee
where in the right-hand side a circular shift on $P$ was done. This
equation is fulfilled if we require that
\be
a(P_1,P_2,...,P_n)\phi _{P_1}(m,0)=
z_{m+1}^{ \f{1-(-1)^{\kappa} }{4} }a(P_2,P_3,...,P_n,P_1) \phi _{P_1}(m-2n,N).
\la{c2}
\ee
Using (\ref{d2}), (\ref{rela}) and (\ref{bb}) we then obtain
\be
\exp \left( i N k_j\right) =-(-1)^{n_{-}}\rho_{m,j}^{(\kappa)}
\prod_{l=1}^nB_{jl}\,,\quad \,j=1,2,...,n  
\la{bethe}
\ee
where
\be
n_{-}=\mbox{int} [ \f{1}{2}\sum_i \delta_{\epsilon_i,-1}  ]
\hspace{.35cm} \mbox{and} \hspace{.35cm} \rho_{m,j}^{(\kappa)}=
\left( \f{ \Omega_{j,m}^2 }{ z_{m+1} } \right)^{\f{1-(-1)^{\kappa}}{4}}.
\la{cons}
\ee
Actually we have in (\ref{bethe}) two distinct sets of equations, a first 
 one for $m=1$ and $m=5$ and a second one  for $m=3$ and $m=7$. These sets
must be solved simultaneously for the validity of the Bethe-ansatz  
(\ref{ansatz}). From Eqs. (\ref{t}) and (\ref{eqqq})-(\ref{omega}) 
we verify that those
equations degenerate ($\rho_{m,j}^{(\kappa)}=1$, $m=1,3,5,7; j=1,2,...,n$) 
in the following cases: (a) $\kappa=0$ or 2 for 
arbitrary values of temperatures $t$, (b) $\kappa=1$ or 3 only at 
the critical temperature $t=t_c=1$. In both cases the Bethe-ansatz
equations are given by
\be
\exp (-iN k_j) =-(-1)^{n_-}\sqrt{\epsilon }%
\prod_{l=1}^nB_{jl}\,,\quad \,\epsilon =\pm 1  
\la{bethet1}
\ee
with $B_{j,l}$ and $n_-$ given by (\ref{bb}) and (\ref{cons}), respectively.
For a given value of $\epsilon=\pm1$ the prefactor $\delta_-=(-1)^{n_-}$,
in Eq. (\ref{bethet1}), may be positive or negative depending on the particular
choice of the set $\{\epsilon_1,...,\epsilon_n\}$. If $n$ is even
these two choices are equivalent. The solution of both equations are
the same except that one of the quasi-moments differs by the value $\pi$
 (mod $2\pi$). However if $n$ is odd the situation is different, and we 
should consider both equations. Their solution gives us independent 
wave-functions.
\section{The operator content of the  $\,$Baxter-Wu model and the $SCP$}
In this section, by exploiting the conformal invariance at the critical 
point, we are going to derive the operator content of the Baxter-Wu model 
and the $SCP$. The conformal anomaly $c$ and
anomalous dimensions $(x_1,x_2,...)$ are obtained in a standard way
 from the finite-size behaviour of the eigenspectra of the associated
transfer matrix, at the critical temperature. If we write $T=\exp(-\hat
H)$, then in a strip of width $L$ with periodic boundary conditions the
ground-state energy, $E_0(L)$, of $\hat H$ behaves for large $L$ as
\cite{anomalia}
\be
\frac{E_{0}(L)}{L}= \epsilon_{\infty} - \frac{\pi c v_{s}}{6L^{2}} +o(L^{-2}),
\la{ano}
\ee
where $\epsilon_{\infty}$ is the ground-state energy, per site, in the
bulk limit. Moreover, for  each operator $O_{\alpha}$ with dimension
$x_{\alpha}$ there exists a tower of states in the spectrum of $\hat H$
with eigenenergies given by \cite{{cardrev},{cardim}}
\be 
E_{m,m'}^{\alpha}(L)=E_{0}+\frac{ 2\pi v_s }{ L }(x_{\alpha}+m+m') 
+ o(L^{-1}), \\
\la{dim}
\ee
where $m,m'=0,1,2,\ldots$ . 
The factor $v_{s}$ appearing in the last two equations  is the
sound velocity and has unit value  for isotropic square lattices. 
 We can calculate directly the higher
eigenvalues of $\hat T^{BW}_{(a,b)}$ and $T^{SCP}_{(\kappa)}$
of our model by a numerical
diagonalization, using for example the power method.  However since these
matrices are not sparse and have dimension
$2^{L}$, for a horizontal width L,  we cannot compute the eigenspectra
by numerical  diagonalization methods for
$L > L_0 \sim 18$ in the case of the Baxter-Wu model.  

Instead of a direct calculation we can explore the relation (\ref{pbwscp})
 among the
Baxter-Wu model and the $SCP$, and solve numerically the associated Bethe 
ansatz equations derived in section 3. If we write
 $T=\exp(-\hat H)$ for both models, the relation 
(13) implies 
\be
Tr(e^{-M H_{L}^{BW}})=Tr(e^{-M H_{N}^{SCP}}),
\la{tr}
\ee
where $N=\f{2L}{3}$. Its is important to observe that 
although $H_{L}^{BW}$ and $H_{N}^{SCP}$
have the same dimension $2^{L}$ they may have different eigenvalues. 
Indeed that is the case, specially if $t\neq t_c=1$ as we
can verify by a brute force diagonalization of these transfer 
matrices on small lattices. In order to present our results let us
rewrite the Bethe-ansatz equations (\ref{bethet1}) in a convenient form.
 We are going to choose the prefactor $\delta_-=(-1)^{n_-}=1$, since  as we
discussed in the last section, if $n$ is even all the energies can be
obtained from a given choice of $\delta_-$, and for $n$ odd the energies
 obtained by different choices of $\delta_-$ are complex-conjugate pairs.
 The eigenvalues $E_{n}^{\{s_j\}}$ of
$H_N^{SCP}$ in the sector with  $n$ dislocations are given by
\be
 E_{n}^{ \{s_i\} } = -\frac{N}{4}\ln (16t^2(1 + t^2)) - 
\sum_{j=1}^n(e_j^{(s_j)}-ik_j^{(s_j)}),
\la{ener}
\ee
where
\be
 e_j^{(s_j)}=1/2\ln\left( x_j+s_j\sqrt{x_j^2-1}\right), \; \; 
x_j=\cos(2k_j^
{(s_j)})+t+1/t ,
\la{sinal}
\ee
with $1=s_1=s_2=\ldots=s_{n-l}=-s_{n-l+1}=\ldots=-s_{n}$, and
$l=0,1,...,n$ fixed. The quasi-momenta $\{k_j^{(s_j)}\}$ are obtained by
solving the equations 
\be
\exp \left( i N k_j^{(s_j)}\right) =-\sqrt{\epsilon}\prod_{p=1}^n
 \left(\frac{\cosh (e_j^{(s_j)}+ik_p^{(s_p)})}{\cosh (e_p^{(s_p)}+
ik_j^{(s_j)})}\right); \;\; j=1,2,...,n ,
\label{eqb}
\ee
where $\epsilon=1$ ($\epsilon=-1$) gives the even (odd) part of the 
eigenspectrum, with respect to the $Z(2)$ symmetry of the $SCP$ discussed
in the last section. Strictly speaking these energies may only give part of
the eigenspectra, since the completeness of the Bethe-ansatz solution
presented in the last section is an open question. Numerically we have studied 
these equations extensively for lattice sizes up to $N\sim 200$ and
part of our results at $t=t_c=1$ were presented in \cite{ax}. For
example, the ground-state energy for $H_N^{SCP}$ corresponding to the
boundary conditions $\kappa=0,1,2$ and 3 given in (\ref{cor}) belongs to the sector
where $n=N-\kappa$, $s_1=s_2=\cdots=s_n=1$ and $\epsilon=1$. In table 1 we
present these ground-state energies, per site, for $\kappa=0,1$ and 2 
($\kappa=3$  is degenerate with $\kappa=1$).
The conformal anomaly is obtained by using (\ref{ano}). The bulk energy
$\epsilon_{\infty}^{SCP}=-\f{3}{4}\ln 6$ can be obtained from the exact 
solution in the bulk limit \cite{bw} and the sound velocity
 $v_s^{SCP}=\sqrt{3}/3$, can be inferred from (\ref{dim}) and an 
overall analysis of the dimensions  appearing in the
model. With these values the ground-state energy (first column in table 1)
 gives us the estimators $c(N)$ presented in table 2.  
As expected the conformal
anomaly is $c=1$, as for  the 4-state Potts model. The direct calculation of
the eigenspectra of $H^{SCP}$ and $H^{BW}$ for small chains shows us 
that although the eigenspectra of both models is not the same, several
eigenvalues coincide. This is the case for the ground-state.  
Consequently by using the bulk limit value \cite{bw} 
$\epsilon_{\infty}^{BW}=-\f{1}{2}\ln 6$, and the sound velocity
 $v_s^{BW}=\sqrt{3}/2$,  we obtain the expected value $c=1$ for
the Baxter-Wu model.

The dimensions defining the operator content of the model are
obtained from the large-N behaviour of the energies of excited states.
Let us concentrate on the $SCP$. The Bethe-ansatz equations (\ref{eqb})
 are the same
for all boundary conditions, specified by  $\kappa$ (0,1,2 and 3) in (\ref{cor}), only
the allowed values of $n$, given by (\ref{dis}) depend on the particular value
of $\kappa$. Using (\ref{dim}),  the finite-size sequences   for 
some dimensions  are shown in table 3. 
In this table $x_j^{\epsilon}(N-n,l)$
are the estimators of the dimensions associated to the
$j$th lowest eigenenergy in the eigensector labelled by $n$ and $\epsilon$. 
The values of $l$ used in  (\ref{ener})-(\ref{eqb}) to obtain the corresponding
energies are also shown. The numerical solution of (\ref{ener})-(\ref{eqb}) was done by the
Newton type  method. The roots $\{k_j\}$ for all the solutions that we obtained are real. 
Although we cannot discard this possibility for higher states, we do not 
find any string-type solution. The numerical solution of (\ref{ener})-(\ref{eqb}) is not
easy  for $N\sim 200$ due to numerical instabilities, and some tricks are
necessary. In most cases  we solve initially these equations for small
values of $t$, where a good guess can be given, and use the solution 
obtained as the initial guess for a larger value of $t$. We repeat 
this process up to $t=t_c=1$. As a result of our extensive calculation of the
eigenspectra of $H^{SCP}$ we arrive at the following conjecture. Namely, the
dimensions of primary operators in a given sector labelled by $n=N+\kappa+4j$ 
$(j=0,\pm 1, \pm2,...)$  of the $H^{SCP}$ with boundary condition $\kappa$ 
(0,1,2 or 3)  are given by 
\be
x_{p,q}=\f{1}{2}(4p^2+\frac {q^2}{4}), 
\; \; \; p=j-\f{\kappa}{4},\; \; \; \; q=0,\pm 1, \pm 2,... \; .
\la{xp}
\ee
In obtaining (\ref{xp}), the value of $E_0$ in (\ref{dim}) is the value of
the ground-state energy for the periodic ($\kappa=0$) $SCP$. Moreover
the number of descendants with dimensions $x_{p,q}^{(\kappa)}+m+m'$ 
($m,m \in Z$) is given by the product of two independent
Kac-Moody characters. This allows, by using (\ref{ano}) and (\ref{dim}), to
write the generating function $Z_{N\times M}^{SCP}(\kappa)$ of the $SCP$ 
with boundary condition $\kappa$, up to order $\exp({-M/N})$ 
($M,N \rightarrow \infty$) as
\be
Z_{N\times M}^{SCP}(\kappa)=\exp(-e_{\infty}^{SCP}MN)z^{1/12}\Theta^2(z)
\sum_{p\in(Z-\kappa/4)}\sum_{q\in Z}
z^{ \f 12 (4p^2+\f 14 q^2)}
\la{ppscp}
\ee
where 
\be
z=\exp( \f{2\pi N}{M} v_s^{SCP}) \hspace{1.5cm}
 \Theta(z)=\prod_{n=0}(1-z^n)^{-1}.
\la{qt}
\ee
Since the Bethe-ansatz roots $\{k_j\}$ of (\ref{eqb}) are real numbers, we
can apply analytical methods [21-24] based on the 
Wiener-Hopf method to obtain the finite-size corrections of the
 eigenenergies. We calculate the finite-size corrections of the lowest 
 eigenenergies in the sector with $n$ dislocations and parity $\epsilon$.
Since these calculations are rather technical we present them in appendix A 
for the interested reader. These analytical results are in agreement with the
 conjecture (\ref{xp}), 
obtained from the numerical solutions of (\ref{ener})-(\ref{eqb}).

Let us return to the Baxter-Wu model. Consider initially the model 
with periodic boundary conditions ($a=b=1$ in (\ref{sigma})). Comparing the
eigenspectra of $\hat T_{+,+}^{BW}$ and  $\hat T_{(0)}^{SCP}$,
 obtained by a direct diagonalization 
on small lattices,  we verify that many of the
dimensions $x_{p,q}^{(0)}$ appearing in (\ref{xp}) are absent. 
For example,  the energies producing the estimator $x_4^+(0,1)$ in the 5th 
 row of table 3 only appear in $T^{SCP}_{(0)}$. 
Following for large lattices the energies which
are exactly related in both models we verified that the lower dimensions
in the Baxter-Wu model, with periodic boundary condition, 
 are given by $x= 0,\frac{1}{8}, \frac{1}{2}, 
\frac{9}{8}, \ldots $, and appear
with degeneracy $d_{x}=1,3,1,9,...$, respectively. 
Due to its $D(4)$ symmetry the Baxter-Wu model has the same eigenspectra
for the non-periodic boundary  conditions given in (\ref{sigma}), i.e., 
$(a,b)\neq (+,+)$. We have shown, at the end of section 3, that the partition 
function in these cases is exactly related with a 
$SCP$ with boundary  conditions  not included in (\ref{cor}). 
Actually the application of the Bethe-ansatz 
in this case, if possible, is more difficult since the  number of colours 
in a row is not a good quantum number anymore. 
However at $t=t_c=1$ our direct calculations of the eigenspectra on small
 chains shows that there exist exact coincidences between the eigenvalues of 
$\hat T_{(-,-)}^{BW}$ and those of $\hat T_{(\kappa)}^{SCP}$, which are given 
by the Bethe-ansatz equations (\ref{ener})-(\ref{eqb}). These coincidences 
enable us to 
verify that the lower dimensions of the Baxter-Wu model with boundary
condition $(a,b)\neq(+,+)$ in (\ref{sigma}) are given by $x=1/8,1/2,10/16,...$ 
and appear with 
degeneracy $d_x=1,1,4,...$  These are the same dimensions reported in \cite{baake}
for the 4-state Potts model with antiperiodic boundary condition.
These results supplemented with the global eigenspectrum calculated for 
small systems, indicate that the operator content of the Baxter-Wu model is
the same as that of the 4-state Potts model \cite{baake} and is given in 
terms of a $Z(2)$ orbifold \cite{ginsparg} of the Gaussian model.

Before closing this section, since we have calculated the eigenspectra
of $H^{BW}$ and $H^{SCP}$ for large lattices we can also calculate the
dimensions of the operators responsible for the corrections to  finite-size
scaling in both models. Since these calculations were already presented 
earlier (see Eqs.(11)-(12) and Table 3 in \cite{ax}), we only mention that 
$x_\gamma=4$ is the lowest dimension of the operator responsible for the 
finite-size  
deviations of the critical behaviour. This means that relations (\ref{ano}) and
(\ref{dim}) have corrections which are power-like with the system size $L$. 
These corrections are like those of the Ising model and different from
those of the 4-state Potts model. 
This explains why the finite-size studies of the Baxter-Wu model
 have good convergence, in contrast with the 4-state Potts model, where 
the operator responsible for these corrections is marginal $(x_\gamma=2)$
producing logarithmic behaviour with the system size.
\section{The off-critical properties of the Baxter-Wu model}
The Baxter-Wu model and the $SCP$ have a massive spectra at $t\neq t_c=1$. 
A continuum field theory describing the long-distance physics in this
phase can be obtained in the neighborhood of the perturbing thermal 
parameter $\delta=t-t_c \ns 0$. Such a field theory will be massive and 
the masses can be estimated from the finite-size behaviour of the 
eigenspectra of $H=-\ln \hat T$. We can calculate the mass spectrum by
applying the scheme proposed by Sagdeev and Zamolodchikov \cite{sz}
in the study of the Ising model under the influence of magnetic perturbations. 
According to this scheme we should initially calculate the finite-size
 corrections of the zero-momenta eigenenergies $E_k(\delta,L), \hspace{.25cm}k=0,1,2,...$, 
at the conformal invariant critical point $\delta=0$. From our
analysis presented in the last section these corrections are governed mainly 
by an irrelevant operator with dimension $x_\gamma=4$ and have integer 
power-law behaviour with the system size $L$. According to 
conformal invariance \cite{{cardim}} 
$E_k(\delta,L)$ should behave as
\be
E_k(L)=e_\infty L +\f{2\pi v_s}{L}\left( x_k-\f{c}{12} \right)
+ a_{k,1}L^{-3}+a_{k,2}L^{-5}+\cdots ,
\la{ee}
\ee
where $x_k$ is the conformal dimension associated to $E_k$ and $a_{k,i}$ 
($i=1,2,...$) are $L$-independent factors. According to the scheme of 
\cite{sz}, if the perturbed operator which  produces the massive 
behaviour has dimension $y$, we should calculate the eigenspectra
in the asymptotic regime $\delta\rightarrow0$, $L\rightarrow\infty$, with
\be
X=\delta^{\f{1}{2-y}}L
\la{sr}
\ee
kept fixed. In this regime (\ref{ee}) is replaced by
\be
E_k(\delta,L)=e_\infty L +\delta^{\f{1}{2-y}}F_k(X)
+ a_{k,1}\delta^{\f{3}{2-y}}G_k(X)+a_{k,2}\delta^{\f{5}{2-y}}H_k(X)+\cdots\,\, .
\la{ee2}
\ee
The masses of the continuum field theory are obtained from
the large-$X$ behaviour of the functions \cite{sz} $F_k(X)$, and 
are given by
\be
m_k\sim F_k(X)-F_0(X),
\la{mm}
\ee
where $F_0(X)$  is associated  in (\ref{ee2}) with the ground-state energy.

In the present application, the thermal fluctuations  are produced by the
energy operator, which has dimension $y=x_{\epsilon}=1/2$. Since we are
 going to calculate the eigenenergies of the Baxter-Wu model by exploiting 
its connection with those of the $SCP$, it is important to compare their 
eigenspectra for small lattice sizes. In table 4 we represent for $t<t_c$ $(T>T_c)$ 
the relative 
location in the eigenspectra of the lower zero-momentum energies of both models. The 
eigenenergies in the same line are exactly degenerate on  the finite lattice. 
In this table we also show  the  parity-quantum number 
$\epsilon\pm1$, of the eigenenergies of $H^{SCP}$.

It is important to mention that although the Baxter-Wu model and the
$SCP$ are exactly related, the parameter $t$ has quite a different effect in 
both models. In the case of the Baxter-Wu model it drives the system from an 
ordered phase $(T<T_c,\; t>t_c=1)$ to a disordered phase $(T>T_c,\; t<t_c=1)$.
On the other hand, as we can see form (14), in the $SCP$ it drives the model 
from an ordered phase rich in colours 4 and 8 ($t>t_c=1$) to another ordered
phase rich in colours 2 and 6 $(t<t_c=1)$.   This fact implies that even for 
$T>T_c$ we should have in the $SCP$ an infinite set of states, including the 
ground state, that degenerate exponentially with the system size 
$(E_n-E_n'\sim\exp(-aL))$, which certainly is not the case for the 
Baxter-Wu model in its disordered phase. In table 4 the pair of levels 
$E_i^{SCP}$ and $E_{i+1}^{SCP}$ ($i=0,2$ and 4) degenerate exponentially.
Baxter and Wu \cite{bw} in their original calculation of the exponent
 $\alpha$ of the Baxter-Wu model used the excited energy 
$E_2^{SCP}(\epsilon=1)$, instead of  $E_3^{SCP}(\epsilon=-1)=E_1^{BW}$. 
However, as we mentioned, these energies become exponentially degenerate with
system size, not changing  their exact result $\alpha=2/3$.

Exploring the correspondences presented in table 4
 and using (\ref{sr})-(\ref{mm}) we can
calculate the mass ratios of the underlying massive field theory governing
the Baxter-Wu model for $T\neq T_c$. They are calculated from the 
asymptotic regime $X\rightarrow\infty$  of the finite-size sequences
\be
R_k(X,L)=\f{F_k(X,L)-F_0(X,L)}{F_1(X,L)-F_0(X,L)}\rightarrow \f{m_k}{m_1}.
\la{smm}
\ee
The functions $F_k(X,L)$ are obtained by using in (\ref{ee2}) the finite-size 
sequences of the zero-momentum states ($k$=0,1,2...). The exact degeneracy
of $E_1^{BW}$ (see table 4) implies the first mass $m_1$ is triple generated.
 From the equality $E_2^{BW}=E_4^{SCP}$ we can calculate the second mass
$m_2$ by solving the Bethe-ansatz equations derived in section 3 for the 
$SCP$. Unfortunately, although trying hard, we were not able to find the 
Bethe-ansatz roots that would correspond to this energy. However, applying
the Lanczos method directly in $H^{SCP}$ we calculate this eigenenergy up
to $L=21$, in the Baxter-Wu model. In table 5 we show the estimators 
$R_2(X,L)$ obtained by using in (\ref{ee2})-(\ref{smm}) $L=15,18$ and 21. 
These results are
consistent with the conjecture $m_2=\sqrt{3}m_1$. In the case of mass $m_3$ our
results are more precise since we were able to calculate 
$E_3^{BW}=E_6^{SCP}(\epsilon=1)$ for lattice sizes up to $L$=150, by 
solving the Bethe-ansatz equations (\ref{ener})-(\ref{eqb}). In table 6 we
 show the estimator $R_3(X,L)$ for some values of $X$, obtained by
using in (\ref{ee2})-(\ref{smm}) $L=144,147$ and L=150.
These results show clearly that $m_3=2m_1$. The numerical analysis of other
higher energies in the spectrum shows that a continuum starts at $m_3$.

The mass ratios we obtained should be the same as those of the 4-state Potts model, 
since we expect both models share the same universality class 
of critical behaviour. In fact they coincide with the masses previously 
conjectured \cite{mpter} for the 4-state Potts model and are also given by
the  masses of a sine-Gordon model \cite{sg} at a special coupling, i. e.,
\be
m_{i+1}=m_1\sin(\f{\pi}{6}i) , \hspace{.25cm}i=1,2,3.
\la{mt}
\ee

To conclude this section we mention that we also studied the effect of
magnetic perturbations in the Baxter-Wu model. This was done by calculation 
of the eigenspectra of $\hat T_{L,M}^{BW}$ with the addition of an external 
magnetic field $h$. In this case $\delta=h$ and $y=1/8$ in (\ref{ee2}), and
the masses we obtained are consistent with those reported in \cite{mpmag}
 for the 4-state Potts model. However our results in this case, specially for 
larger masses, lack precision because we had to calculate directly the
eigenspectrum of $\hat T_{L,M}^{BW}$, since the equivalence with the
$SCP$ presented in section 2 is not valid anymore and unlike $\hat T_{N,M}^{SCP}$ 
this matrix is not sparse.
\section{Conclusions and Comments}
The operator content of the Baxter-Wu model was calculated for several
 boundary conditions by exploiting the conformal invariance of the 
infinite system at the critical point. Our results are calculated 
analytically and numerically for very large lattice sizes. This was possible 
due to the relation between the Baxter-Wu model and the $SCP$. Actually 
we showed that the partition functions of both models are exactly related for 
several boundary conditions (see section 2) and we were able to extend the
 original Bethe-ansatz solution \cite{bw} for most of these boundaries 
(see section 3)

The operator content of the $SCP$ with several toroidal boundary conditions 
(see (\ref{xp})-(\ref{qt})) is the same as those of a Gaussian model with 
dimensions \cite{kada}
\be
 x_{n,m}=gn^2+\f{m^2}{4g},
\la{xg}
\ee
where $g=1/8$ is the compactification  radius and $n,m\in Z$ are the 
vorticity and spin-wave number, respectively. However only part of the 
eigenspectra of both models coincide. Our analysis (section 4) shows that 
the dimensions of the Baxter-Wu model, for several boundary conditions are 
given by a $Z(2)$ orbifold of the above Gaussian model. This operator content 
coincides with the 4-state Potts model, indicating that indeed
 both models share the same universality class of critical behaviour. It is 
interesting to remark that whereas for the $SCP$ the operator content is given 
in terms of characters of the Kac-Moody algebra, in the Baxter-Wu model the 
characters are those of the Virasoro algebra.

On the other hand a similar exact relation as that between the $SCP$ and the
 Baxter-Wu model also exists between the 4-state Potts model  and the 
6-vertex model at its isotropic point ($\gamma=0$), or equivalently the 
quantum XXX chain (anisotropy $\gamma=0$). The operator content of this 
model is given by a combination of the dimensions given in (\ref{xg}) but with
 $g=1/2$. In the 6-vertex model, or XXZ chain, the point where $g=1/8$
 corresponds to the anisotropy $\gamma=3\pi/4$, and is the so-called
 Kosterliz-Thouless point. This implies that exactly at the critical point
 ($L\rightarrow\infty, T=T_c$) the Baxter-Wu model
 and the 4-state Potts model are governed by the same conformal theory, 
but deviations from the critical point, like for example the finiteness 
of the lattice, will be governed by an effective Gaussian model with
different compactification radius. It is known \cite{alcbar} that in the 
case of the 6-vertex model or XXZ chain the finite-size corrections are 
ruled mainly by the operator with dimension $x_{0,2}$ in (\ref{xg}) besides the
 descendant of identity operator with dimension $4$. 
This implies the appearance
 of logarithmic corrections, with the system size, at $g=1/2$ since 
$x_{0,2}=2$  and the corresponding operator responsable for such corrections 
is  marginal. 
On the other hand at $g=1/8$ we only  have integer power-law
 corrections with the system size, since in this case the operator with 
dimension 4  dominates the finite-size correction. 
This explains why although the Baxter-Wu 
model and the 4-state Potts model are given by the same $Z(2)$ orbifold of a
 Gaussian theory, they show quite different behaviour at finite lattices.

In section 5 we calculated the mass spectrum of the underlying field theory
 governing the Baxter-Wu model around its critical point. In the case of 
thermal perturbations we obtained the masses given in (\ref{mt}) which are 
the same as those of the 4-state Potts model \cite{mpter} and are also the
 masses of a special point of a massive sine-Gordon field theory \cite{sg}.
 Finally in the case of magnetic perturbation our numerical results, although 
poorer, are consistent with the same masses reported earlier for the 4-state
 Potts model \cite{mpmag}.

\begin{center}
{\bf Acknowledgments}
\end{center}

It is a pleasure to acknowledge profitable discussions with M. J. Martins
and  M. T. Batchelor for a careful reading of 
our manuscript. This work was supported in part by Brazilian agencies
 CNPq, FAPESP and FINEP.

\newpage

\renewcommand{\theequation}{A.\arabic{equation}}
\setcounter{equation}{0}

\begin{center}
{\bf \Large Appendix}
\end{center}
\appendix{}
\section{Analytic calculation of the leading finite-size corrections}
In this appendix we calculate analytically the leading finite-size corrections 
for some of the  eigenenergies of the $SCP$ at the critical point $t=t_c=1$. 
We will calculate the finite-size corrections of the lowest energies $E_n(\epsilon$) 
of the Hamiltonian, $H^{SCP}=-\ln\hat T_{SCP}$, in the sector with $n$ dislocations and
$Z(2)$-colour parity $\epsilon$ $(n=0,1,2,...;\;\epsilon=\pm1)$. Since the associated 
roots of the Bethe-ansatz equations are real numbers our analytical calculations are based on
the method pioneered by de Vega  and Woynarovich \cite{corr1} and  Hamer \cite{corr2} and 
refined by Woynarovich and Eckle \cite{corr3} (see also \cite{corr4}). In order to apply this
method it is convenient to change the variables 
$\{ k_1,k_2,...,k_n\}$ appearing in (\ref{bethet1}) into new variables 
$\{u_1,u_2,...,u_n\}$ so that $B_{j,l}$ become a function of the difference $u_j-u_l$. 
This was
done by Baxter and Wu \cite{bw} for arbitrary temperatures  and $B_{j,l}(u_j-u_l)$ 
are now given in terms of elliptic functions. At
$t=t_c=1$ these elliptic functions become hyperbolic functions, with
\be
B_{j,l}=-\exp\left( -i\Theta(u_j-u_l) \right)=-i\tanh(u_l-u_j-i\pi/4).
\la{eq1}
\ee
In terms of the variables $u$ the quasimomenta $k(u)$ and the factors $e(u)$ 
in (\ref{sinal}) are given by
\be
k(u)=i/2\ln
\left( 
\f{ \tanh(i\pi/8+u)}{\tanh(i\pi/8-u)}
\right),
\la{k}
\ee
\be
e(u)=1/2\ln
\left(
\f{\cosh(2u)+\sqrt{2}/2}{\cosh(2u)-\sqrt{2}/2}
\right).
\la{e}
\ee
The Bethe-ansatz equations (\ref{bethet1}) or (\ref{eqb}) can be written as
\be
\f{I_j}{N}=\f{1}{2\pi}
\left\{ k(u_j)+\f{1}{N}\sum_{l=1}^{n}
\Theta(u_j-u_l)\right\}\hspace{.5cm}j=1,2,...n,
\la{eq.11}
\ee
where $2I_j$ are integers or half-odd integers depending on the value of 
 $\epsilon$. The values of $I_j$ for the lowest eigenenergy in the sector 
with  given  values of $n$ and $\epsilon$, which we are interested in, are
\be
I_1,I_2,...,I_n=- \f{n-1-{\tilde\epsilon}}{2},
 \f{n-1-{\tilde\epsilon}}{2}+1,...,
\f{n-1+{\tilde\epsilon}}{2}, \; \;
\tilde\epsilon=\f{1-\epsilon}{4} . 
\la{ei}
\ee
Following a standard procedure \cite{corr1} we  define the density of roots 
\be
\sigma_N^{n}(u)=\f{dZ_N^{n}}{du},
\la{sn}
\ee
where
\be
Z_N^{n}(u)=
\f{1}{2\pi}
\left\{ k(u)+\f{1}{N}\sum_{l=1}^{n}
\Theta(u-u_l)\right\}.
\la{sn2}
\ee
When $N\rightarrow\infty$ (\ref{sn})  becomes an integral equation whose 
solution gives  the bulk limit of the density of roots. In particular, 
in the sector $n=N$ this density of roots is given by
\be
\sigma_\infty(u)=\f{4}{\pi}
\f{\cosh(4u)\cosh(8u/3)}{\cosh(8u)+1}.
\la{s}
\ee
In this limit the energy per site is given by \cite{bw}
\be
e_\infty^{SCP}=-\f{3}{4}\ln(6).
\la{ein}
\ee
The difference between the energy per site and the density of roots and their 
bulk-limit values can be expressed by
\be
\f{E_N^{n}}{N}-e_\infty^{SCP}
=-\int_{-\infty}^{\infty}f(v)S(v)dv,
\la{ei2}
\ee
and
\be
\sigma_N^{n}(u)-\sigma_\infty(u)=-\f{1}{2\pi}
\int_{-\infty}^{\infty}p(u-v)S(v)dv,
\la{ei1}
\ee
respectively, where
\be
S(v)=\f{1}{N}\sum_{j=1}^{n}\delta(v-u_j)-\sigma_N^{n}(v),
\la{S}
\ee
\be
p(u)=-\f{8\sqrt3}{3}\f{ \sinh(8u/3) } { \sinh(4u) },
\la{pu}
\ee
\be
f(u)=1/2\int_{-\infty}^{\infty}
\f{ \sinh(\pi x/8) } { x\left( \cosh(\pi x/4) -1/2\right) }\exp(ixu)dx.
\la{fu}
\ee
Using the Euler-Maclaurin formula we can expand (\ref{ei2}) and (\ref{ei1}),
 obtaining
\begin{eqnarray}
\f{E_N^{n}}{N}-e_\infty^{SCP}=\left( \int^{\infty}_{\Lambda_+}
f(v)\sigma_N^{n}(v)dv
-\f{1}{2N}f(\Lambda_+)
-\f{1}{12N^2}
 \f{f'(\Lambda_+)}{\sigma_N^{n}(\Lambda_+)}\right)
  \nonumber \\ 
+\left(\int^{\infty}_{\Lambda_-}
\f{ f(v)}{2\pi}\sigma_N^{n}(v)dv-\f{1}{2N}f(\Lambda_-)
-\f{1}{12N^2}
 \f{f'(\Lambda_-)}{\sigma_N^{n}(\Lambda_-)}\right),
\la{wh1} 
\end{eqnarray}
and
{\small
$$
\sigma_N^{n}(u)-\sigma_\infty(u)=\left( \int^{\infty}_{\Lambda_+}
\f{ p(u- v)}{2\pi}\sigma_N^{n}(v)dv-
\f{1}{2N}p(u-\Lambda_+)
+\f{1}{12N^2}
 \f{p'(u-\Lambda_+)}{\sigma_N^{n}(\Lambda_+)} \right)
$$
\be
+\left(\int^{\infty}_{\Lambda_-}
\f{ p(u+ v)}{2\pi}\sigma_N^{n}(v)dv-\f{1}{2N}p(u+\Lambda_-)
-\f{1}{12N^2}
 \f{p'(u+\Lambda_-)}{\sigma_N^{n}(\Lambda_-)}\right),
\la{wh2} 
\ee
}
respectively. In the above equations $\Lambda_+$ and $\Lambda_-$ are the
largest and smallest root determined by the  condition
\be
\int^{\infty}_{ \Lambda_\pm }
\sigma_N^{n}(u)du=
\f{1}{2N}\left( 1+\beta_\pm(n) \right),
\la{sc}
\ee
where 
$$
\beta_\pm(n)=\f{N-n}{2}\mp \f{1-\epsilon}{4} .
$$
We should now consider separately the cases $u>\Lambda_+$ and 
$u<-\Lambda_-$. In the case where $u>\Lambda_+$($u<-\Lambda_-$) the 
corrections of $O(1/N^2)$ are calculated by neglecting the terms in
the second (first), large bracket in (\ref{wh2}). Defining
\be
\begin{array}{lll}
g(u)=p(u)/2\pi, & f^\pm(u)=\sigma_\infty(u+\Lambda_\pm), & 
\chi^\pm(u)=\sigma_N^{n}(u+\Lambda_\pm), \\
\end{array}
\la{def}
\ee
we can write (\ref{wh2}) as 
\be
\chi^\pm(t^\pm)-f^\pm(t^\pm)=\int_0^\infty g(t^\pm-v)\chi^\pm(v)dv
-\f{1}{2N}g(t^\pm)+
\f{1}{12N^2}\f{g'(t^\pm)}{\sigma_N^{n}(\Lambda_\pm)}.
\la{wh}
\ee
This is precisely the standard form of the Wiener-Hopf equation 
(see, for example Morse and Feshbach \cite{morse}). Its solution is
obtained on defining the Fourier transforms
\be
\tilde\chi^\pm_\pm(w)=\int_{-\infty}^\infty 
\exp(iwt)\chi^\pm_\pm(t) dt \hspace{1cm} \hspace{.5cm} 
\chi^\pm_\pm(t)=
\left\{ \begin{array}{ll}
\chi^\pm(t) & t \mn 0\\
    0           & t \nm 0\\
\end{array}\right. 
\la{tf}
\ee
and the corresponding Fourier pairs $ g\leftrightarrow\tilde g$, 
$f\leftrightarrow\tilde f$. Using the fact that
\be
(1-\tilde g(w))^{-1}=G_+(w) G_-(w),
\la{eqgg}
\ee
where
\be
G_+(w)=\f{ \sqrt{2\pi}\Gamma(1/2-iw/4)\exp(iw\ln(2)/4) }
{\Gamma(5/6-iw/8)\Gamma(1/6-iw/8)}=G_-(-w),
\la{GG}
\ee
we can express $\tilde\chi^\pm_\pm(w)$, after some algebraic manipulations as 
\be
\tilde\chi^\pm_+(w)=C^{\pm}(w)+G_+(w)(Q_+^{\pm}+ P^\pm(w)),
\la{eq3}
\ee 
where
\be
C^{\pm}(w)=\f{1}{2N}+\f{iw}{12N^2\sigma_N^{n}(\Lambda_\pm)}
\hspace{.5cm}
Q_+^\pm(w)=\f{2}{\pi}
\f{G_+(i4/3)\exp(-4/3\Lambda_\pm)}{4/3-iw},
\la{cw}
\ee
\be
P^\pm(w)=-\f{1}{2N}+\f{i(g_1-w)}{12N^2\sigma_N^{n}(\Lambda_\pm)}
\hspace{.5cm} g_1=109/6.
\la{pw}
\ee
Equations (\ref{wh}) and the definitions (\ref{def}) give us 
\be
Q^\pm_+(0)=\f{3}{2\pi}G_+(i4/3)\exp(-4/3\Lambda_\pm)=
\f{1}{2N}-\f{ig_1}{12N^2\sigma_N^{n}(\Lambda_\pm)}+
\f{1}{2N}\f{\beta_\pm}{G_+(0)},
\la{eq4}
\ee
\be
\sigma_N^{n}(\Lambda_\pm)=
\f{g_1^2/2-4/3ig_1}{12N^2\sigma_N^{n}(\Lambda_\pm)}+
\f{ig_1+4/3}{2N}+\f{4/3\beta_\pm(n)}{2NG_+(0)}.
\la{ss}
\ee
Finally, using (\ref{eq3}),(\ref{eq4}) and (\ref{ss}) in (\ref{wh1}) and approximating 
$f(|u|)=\sqrt{3}\exp(-4/3|u|)$, $u\gg1$, we obtain the first-order correction for
the lowest energy $E_N^{n,\epsilon}=E_N^{(n)}$ in the sector with $n$ 
dislocations and colour parity $\epsilon$,
\be
\f{E_N^{n}}{N}-e_\infty^{SCP}=\f{\pi v_s^{SCP}}{6N^2}
(-1/6 +2X_r^{\epsilon})+o(1/N^2),
\la{eq6}
\ee
where
 \be
X_n^{\epsilon}=\f{(N-n)^2}{8}+\f{(1-\epsilon)^2}{16}.
\la{dim2}
\ee
In particular for the ground-state energy $n=N$, $\epsilon=1$, we have
\be
\f{E_N^{0}}{N}\equiv\f{E_N^{N,+}}{N}=e_\infty^{SCP}-\f{\pi v_s^{SCP}}{6N^2}
+o(1/N^2),
\la{eqq6}
\ee
and comparing with (\ref{ano}) we obtain the value $c=1$ for the conformal
anomaly. If we now consider  the gaps with respect to the ground-state 
 with periodic boundary  condition we obtain
$$
\f{E_N^{N,\epsilon}}{N}-\f{E_N^{0}}{N}=
\f{ 2\pi v_s^{SCP} X_n^{\epsilon} }{N^2}+o(1/N^2).
$$
Comparing this expression with (\ref{dim}) we obtain the conformal dimensions
$ X_n^{\epsilon}$. These values are in perfect agreement with
the operator content conjectured for the $SCP$ and Baxter-Wu model, presented
in section 4.

%
%
%

%
%
%
%

%
%
%
\newpage
\vspace*{-3cm}
\Large 
\bc 
Figure and Table Captions 
\ec
\normalsize
\noindent Figure 1 - A triangular lattice with $L=3$ rows and $M$=6 columns. 
The Baxter-Wu model is defined on the triangular lattice formed by 
 the points {\large{$\circ $}}, $\Box$   and $\triangle $, and the link
variables $\{\lambda\}$ are defined on the hexagonal lattice formed by
the points $\Box$ and {\large{$\circ $}}.
The site-colouring problem ($SCP$)  is defined on the hexagonal lattice  
formed by the points $\triangle $ and {\large{$\circ $}}. The open
symbols {\large{$\circ $}}, $\Box$   and $\triangle $ denote the
bordering sites related through (2) with the bulk ones (filled simbols).

\vspace{1cm}
\noindent Figure 2 - Examples of configurations for the $SCP$ with
lattice size $N=4$.  The dislocations are represented by  dotted lines.
In $(a)$ the configurations $(m;X)=(1;2,3,4,4)$ and 
$(m';X')=(3;2,2,3,4)$ are in the sector $n=4$ $(\kappa=0)$.
In $(b)$ the configurations $(m;X)=(1;2,3,4)$ and 
$(m';X')=(3;2,2,2)$ belong to the sector $n=3$ $(\kappa=1)$.

\vspace{1cm}

\noindent Table 1 - Ground-state energies per site  for the $SCP$ with lattice
size $N$ and  boundary conditions $\kappa=0,1$ and 2 given by (15). 

\vspace{1cm}

\noindent Table 2 - Conformal anomaly estimators $c(N)$, as a function of 
the lattice size $N$, for the $SCP$ and Baxter-Wu model.

\vspace{1cm}
\noindent Table 3 - Scaling dimensions estimators $x_j^{\epsilon}(N-n,l)$, 
as a function of the lattice size $N$, 
for some eigenenergies. These energies are the $j^{th}$ lowest energy 
obtained  by solving 
 (\ref{ener})-(\ref{eqb}) with values $n$, $\epsilon$ and $l$.

\vspace{1cm}

\noindent Table 4 - Energies  for
the Baxter-Wu model and $SCP$. The energies of the same line are
identical.

\vspace{1cm}

\noindent Table 5 - The mass-ratio estimators $R_2(X,L)$ defined in
(\ref{smm}).

\vspace{1cm}

\noindent Table 6 - The mass-ratio estimators $R_3(X,L)$ defined in
(\ref{smm}).

\newpage
\Large
\bc
Table 1
\ec
\normalsize
%
%
\btb[h]
\bc
\bt{||c|c|c|c||} \hline
 N    &  $\kappa=0$ & $\kappa=1$  & $\kappa=2$  \\ \hline\hline
 6    &-1.3521881950 &  -1.3396181681 & -1.3012024788  \\ \hline
 10   &-1.3468393066 &  -1.3423074850 & -1.3286264987\\ \hline
 50   &-1.3439405169 &  -1.3437591407 & -1.3432148801\\ \hline
 100  &-1.3438498316 &  -1.3438044868 & -1.3436684443\\ \hline
 150  &-1.3438330374 &  -1.3438128841 & -1.3437524226\\ \hline
 200  &-1.3438271593 &  -1.3438158231 & -1.3437818139\\ \hline
\et
\ec
\etb
%
\Large
\bc
Table 2
\ec
\normalsize
%
%
\btb[h]
\bc
\bt{||c|c||} \hline
 N    &  $c(N)$   \\ \hline\hline
 6    &0.996590995    \\ \hline
 10   &0.998910268      \\ \hline
 50   &0.999959561    \\ \hline
 100  &0.999989915    \\ \hline
 150  &0.999995519     \\ \hline
 200  &0.999997480      \\ \hline
\et
\ec
\etb
%

\newpage

%
\Large
\bc
Table 3
\ec
\normalsize
%
%
{\footnotesize
\btb[h]
\bc
\bt{||c||c|c|c|c|c|c||l||} \hline
 N               &  6 & 10  & 50 & 100 & 150 & 200 & \mbox{exact}  \\ \hline
 $x_1^{-}(0,0)$  &0.12502803 & 0.12501702 & 0.12500083 & 0.12500021 & 0.12500009 & 0.12500005 & 0.125  \\ \hline 
 $x_2^{-}(1,0)$  &0.24896741 & 0.24959771 & 0.24998323 & 0.24999580 & 0.24999813 & 0.24999895 & 0.25 \\ \hline
 $x_2^{+}(0,1)$  &0.50626226 & 0.50215317 & 0.50008406 & 0.50002099 & 0.50000933 & 0.50000525 & 0.5  \\ \hline
 $x_1^{-}(2,0)$  &0.62613504 & 0.62548322 & 0.62502093 & 0.62500524 & 0.62500233 & 0.62500131 & 0.625 \\ \hline
 $x_4^{+}(0,1)$  &0.98648357 & 0.99698967 & 0.99992101 & 0.99998057 & 0.99999139 & 0.99999516 & 1  \\ \hline
 $x_1^{+}(3,0)$  &1.16335432 & 1.13836199 & 1.12552520 & 1.12513123 & 1.12505831 & 1.12503280 & 1.125 \\ \hline
 $x_1^{-}(3,0)$  &1.27059213 & 1.25887795 & 1.25038505 & 1.25009649 & 1.25004290 & 1.25002413 & 1.25 \\ \hline
 $x_6^{+}(0,2)$  &1.53502476 & 1.51231634 & 1.50048724 & 1.50012177 & 1.50005411 & 1.50003044 & 1.5  \\ \hline
 $x_{10}^{+}(0,1)$  &1.82581639 & 1.97119336 & 1.99980095 & 1.99995795 & 1.99998195 & 1.99998997 & 2  \\ \hline
\et
\ec
\etb
}
%
%
%

\Large
\bc
Table 4
\ec
\normalsize
%
\btb[h]
\bc
\bt{||c|l||} \hline
 BW &  $SCP$                  \\ \hline\hline
       &  $E_7^{SCP}(\epsilon=1)$      \\ \hline
 $E_3^{BW}$ &  $E_6^{SCP}(\epsilon=-1)$      \\ \hline
       &  $E_5^{SCP}(\epsilon=-1)$      \\ \hline
 $E_2^{BW}$ &  $E_4^{SCP}(\epsilon=1)$      \\ \hline
       &  $E_3^{SCP}(\epsilon=1)$      \\ \hline
 $E_1^{BW}$ &  $E_2^{SCP}(\epsilon=-1)$      \\ \hline
       &  $E_1^{SCP}(\epsilon=-1)$       \\ \hline 
 $E_0^{BW}$ & $E_0^{SCP}(\epsilon=1) $    \\ \hline
\et
\ec
\etb
%

\newpage

\Large
\bc
Table 5
\ec
\normalsize
%

\btb[h]
\bc
\bt{||c|l||} \hline
 X &  $m_2/m_1$                  \\ \hline\hline
 2 &  1.4049      \\ \hline
 5 &  1.7169      \\ \hline
 6 &  1.7273       \\ \hline 
 7 &  1.7303    \\ \hline
\et
\ec
\etb

%

\Large
\bc
Table 6
\ec
\normalsize
%

\btb[h]
\bc
\bt{||c|l||} \hline
 X &  $m_3/m_1$                  \\ \hline\hline
 10 &  2.02632      \\ \hline
 20 &  2.00538      \\ \hline
 40 &  2.00123       \\ \hline 
 40 &  2.00053    \\ \hline
\et
\ec
\etb

%


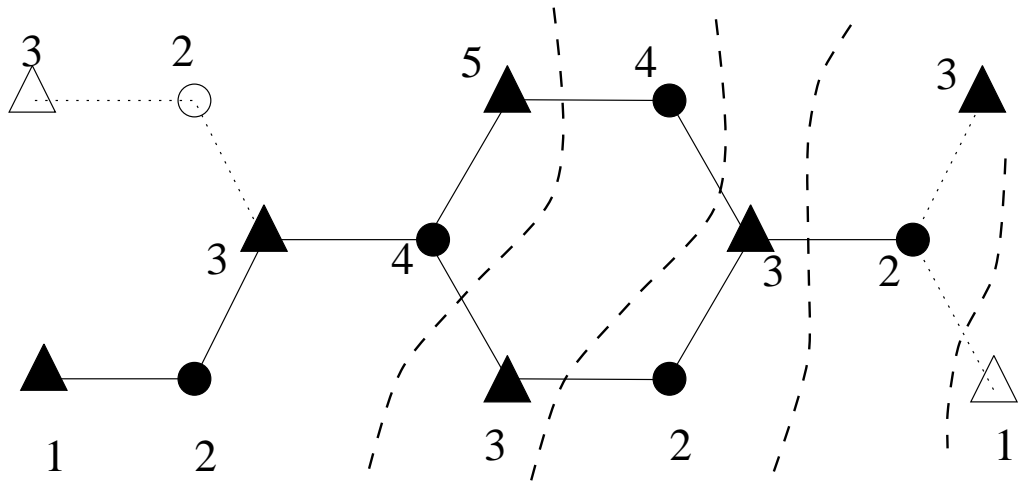
\end{document}